\documentclass{article}
\usepackage[utf8]{inputenc}

\usepackage{natbib}
\usepackage{graphicx}
\usepackage{hyperref}
\usepackage[dvipsnames]{xcolor}
\usepackage[frozencache,cachedir=.]{minted}

\hypersetup{
  colorlinks   = true, 
  urlcolor     = RoyalBlue, 
  linkcolor    = Violet, 
  citecolor    = RoyalPurple 
}

\title{Real-time stream processing in radio astronomy}
\author{Danny Price}
\date{December 2019}

\begin{document}

\maketitle

\section{Introduction}

\begin{figure}
	\begin{center}
	\includegraphics[width=1.0\columnwidth]{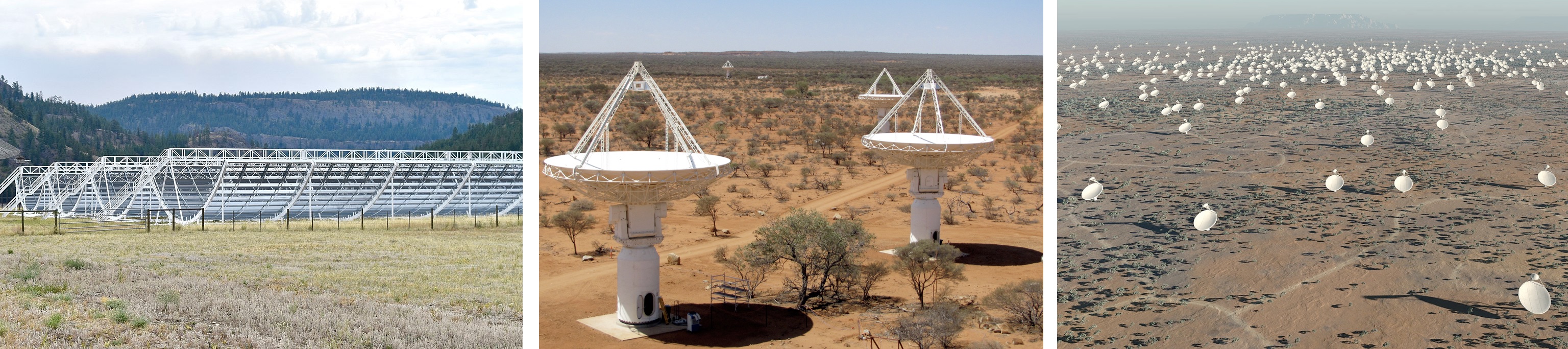}
	\caption{The Canadian Hydrogen Intensity Mapping Experiment (CHIME, left), the Australian Square Kilometre Array Pathfinder (ASKAP, center), and artist's conception of the Square Kilometre Array telescope (SKA, right). Image credits: Z22/Wikimedia, Ant Schinckel, SKA Project Development Office}\label{fig:scopes}
	\end{center}	
\end{figure}

A major challenge in modern radio astronomy is dealing with the massive data volumes generated by wide-bandwidth receivers. For example, in the Square Kilometre Array SKA1-mid telescope, each of 133 antennas is expected to generate 100 Gb/s, for 13.3 Tb/s of data that must be processed in real time \citep{SKA:2015}. These data will be augmented with an extra 64 streams of 40 Gb/s data from the existing MeerKAT telescope, bringing the total data rate to 15.8 Tb/s; within the correlator the aggregate data rate will exceed 57 Tb/s! Overall, the SKA1 is expected to produce over 5 times the global internet traffic as of 2015 \citep{SKA:2016}. Existing telescopes are already producing massive data volumes: for example, the Australian SKA Pathfinder ASKAP, has a pre-beamformer data rate of 103 Tb/s \citep{Askap:2012}, and the Canadian Hydrogen Intensity Mapping Experiment (CHIME) digitizes and processes an impressive 13.1 Tb/s \citep{ChimeFRB:2018}. These telescopes are shown in Figure\,\ref{fig:scopes}.

Such massive data rates are difficult to redistribute, and cannot be recorded as final data products. Data rates are often too great for a single device to cope, and so processing must be split across multiple devices working in parallel. These devices must work in unison to process incoming data in real time, reduce the data volume to a manageable size, and output a science-ready data product.

The aim of this chapter is to give a broad overview of how digital systems for radio telescopes are commonly implemented, with a focus on real-time stream processing over multiple compute devices. Accompanying chapters delve deeper into specific aspects: Chapter 3 gives an overview of pre-processing pipelines on field-programmable gate arrays (FPGAs), and Chapter 5 discusses how signals from a telescope are digitized, split into channels, and prepared for data transport.

In this chapter, we use cross-correlators as a primary example of a stream processing pipeline. We refer readers unfamiliar with correlators to the freely-available \citet{TMS:2017}, which gives a comprehensive overview of interferometry in radio astronomy. 

This chapter is structured as follows. First, we introduce the concept of stream processing (Section\,\ref{sec:stream-processing}), then summarize system architectures and technologies that are commonly used (Section\,\ref{sec:hetero-dsp}). Section\,\ref{sec:ethernet} gives an overview of Ethernet, which has become ubiquitous for data transport between compute nodes. We then discuss data preprocessing (`first-stage' signal processing, Section\,\ref{sec:frontend}), to prepare the data streams for redistribution. Section\,\ref{sec:transport} details logistical concerns of data redistribution, and the common problem of `corner-turning' for data transposition. Section\,\ref{sec:backend} discusses the `second-stage' data processing to form final science data products, and strategies for data recording. We conclude with a discussion of promising technologies for future systems. 

\section{Stream processing}\label{sec:stream-processing}

A telescope's output can be considered as a stream of data; unlike files, a data stream does not have a clearly defined start or endpoint. It is illustrative to think of a telescope's output as a stream of \emph{frames}, where each frame is an array of data at a given time step. A frame may have multiple dimensions, depending to the number of inputs and transformations applied to the data. For example, a frame may have a shape 
\begin{equation}
	(N_{\rm{ant}}, N_{\rm{beam}},N_{\rm{pol}}, N_{\rm{chan}}),
\end{equation}
where $N_{\rm{ant}}$ is the number of antenna elements in the telescope, $N_{\rm{beam}}$ is the number of beams on the sky, $N_{\rm{pol}}$ is how many polarizations the telescope samples, and $N_{\rm{chan}}$ is the number of frequency channels that the data are divided into. For simplicity, we assume all data in the array is of the same datatype (e.g. 8-bit signed integer). 

From this view, the purpose of a telescope's digital signal processing (DSP) system is to apply a series of transformations from one or more frame \emph{sources} to produce a data product that is written to one or more outputs, or \emph{sinks}. One may define a basic set of operations, or \emph{blocks}:

\begin{itemize}
	\item \emph{Stream splitting/merging.} A frame can be split upon an axis to produce multiple data streams with smaller frames. Alternatively, multiple data streams may be merged into a single frame. The rate at which frames are output (i.e. the frame rate) remains the same, but the size of the frame will change, which will affect the output data rate.
	\item \emph{Intra-frame transformation.} Using only data within the frame, a transform is applied to modify the data. The shape and datatype of the frame may change, but the rate at which frames are processed remains constant.
	\item \emph{Inter-frame transformation.} Data from multiple frames are combined; for example, averaged in time to lower frame rate, or buffered up to create an additional frame axis (e.g. a `subframe' axis). Unlike intra-frame transforms, the output frame rate may change.
\end{itemize}

\begin{figure}
	\begin{center}
	\includegraphics[width=1.0\columnwidth]{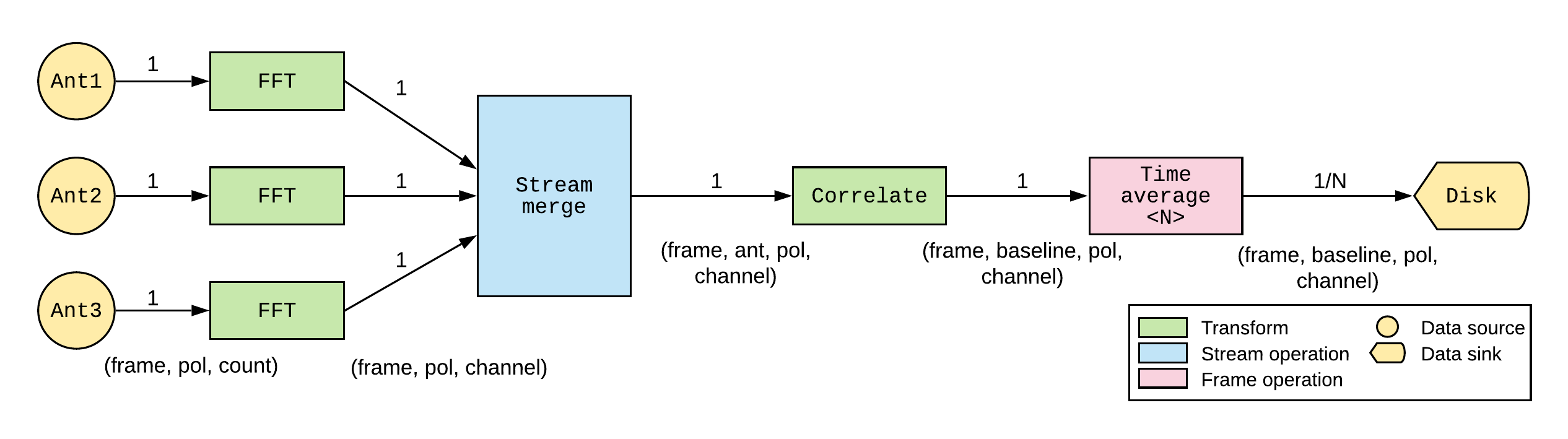}
	\caption{Simple example of a pipeline to cross-correlate three antennas, using data streams. Each antenna outputs a data stream, and a Fast Fourier Transform (FFT) is applied to each to form channels. The three streams are merged together, and then cross-correlation is applied. Multiple frames are then averaged together, before being written to disk. The number above the arrow represents the frame rate; frame dimensions are shown in brackets below the streams.\label{fig:data-stream}}
	\end{center}	
\end{figure}

Stream processing systems may be thought of as a \emph{pipeline} through which data flows---like water---from the source to the sink\footnote{For those with a background in graph theory, a pipeline is a directed acyclic graph where the vertices are processing blocks.}. The time it takes from start to finish is known as the pipeline latency. Importantly, transformations in a pipeline are run concurrently, not applied in serial. If a processing block does not finish processing a frame before the next one arrives, a \emph{bottleneck} will arise, which will slow the entire pipeline. A pipeline may also have multiple data streams that run in parallel, as long as they are independent from one another at that stage of the pipeline.

A simple example diagram representing a pipelined implementation of a cross-correlator is given in Fig.\,\ref{fig:data-stream}. In the diagram, three antennas act as data sources, which output their data streams in parallel. We have labelled their axes `pol' for polarization, and `count' for data from the analog-to-digital converter (ADC). After applying a Fast Fourier Transform (FFT), which adds a `channel' axis, the three parallel streams are then merged into one, which effectively adds an axis to the data frame (labelled `ant' for antenna). The data frame is then fed to a correlation transform, which changes the frame shape to add a `baseline' axis. Time averaging is then applied, which requires multiple frames to be added together. The output is the written to disk (a data sink) as a file. 

\section{Heterogeneous signal processing}\label{sec:hetero-dsp}

Unless all DSP is conducted on a single processing board, data must be redistributed to other parts of the digital system. A common design approach is to have a DSP `frontend', on which first-stage signal processing and reduction is done, that connects via high-speed links to a DSP `backend' that applies second-stage signal processing. The backend ingests data from one or more frontend streams, and reduces the data into final science data products. This division of labor allows different DSP platforms to be used as desired---an approach known as heterogeneous signal processing. 

It is common, particularly for larger systems, for both first-stage and second-stage systems to consist of multiple DSP boards or servers (or more generally, nodes). These nodes need to be interconnected to form a single system. The choice of how to connect boards is dependent upon what data needs to get where. For example, a correlator might consist of $N$ antennas, each of which is connected to frontend board that channelizes the data. At early stages, the data streams are independent, so frontend boards might be located at the antennas. Cross-correlation between antenna data streams could then be performed by a second-stage board that ingests, merges and processes the $N$ data streams in a central facility. As cross correlations are computed on a per-channel basis, one could parallelize the computation across several boards, each processing a subset of channels from all antennas. 

\begin{figure}
	\begin{center}
	\includegraphics[width=1.0\columnwidth]{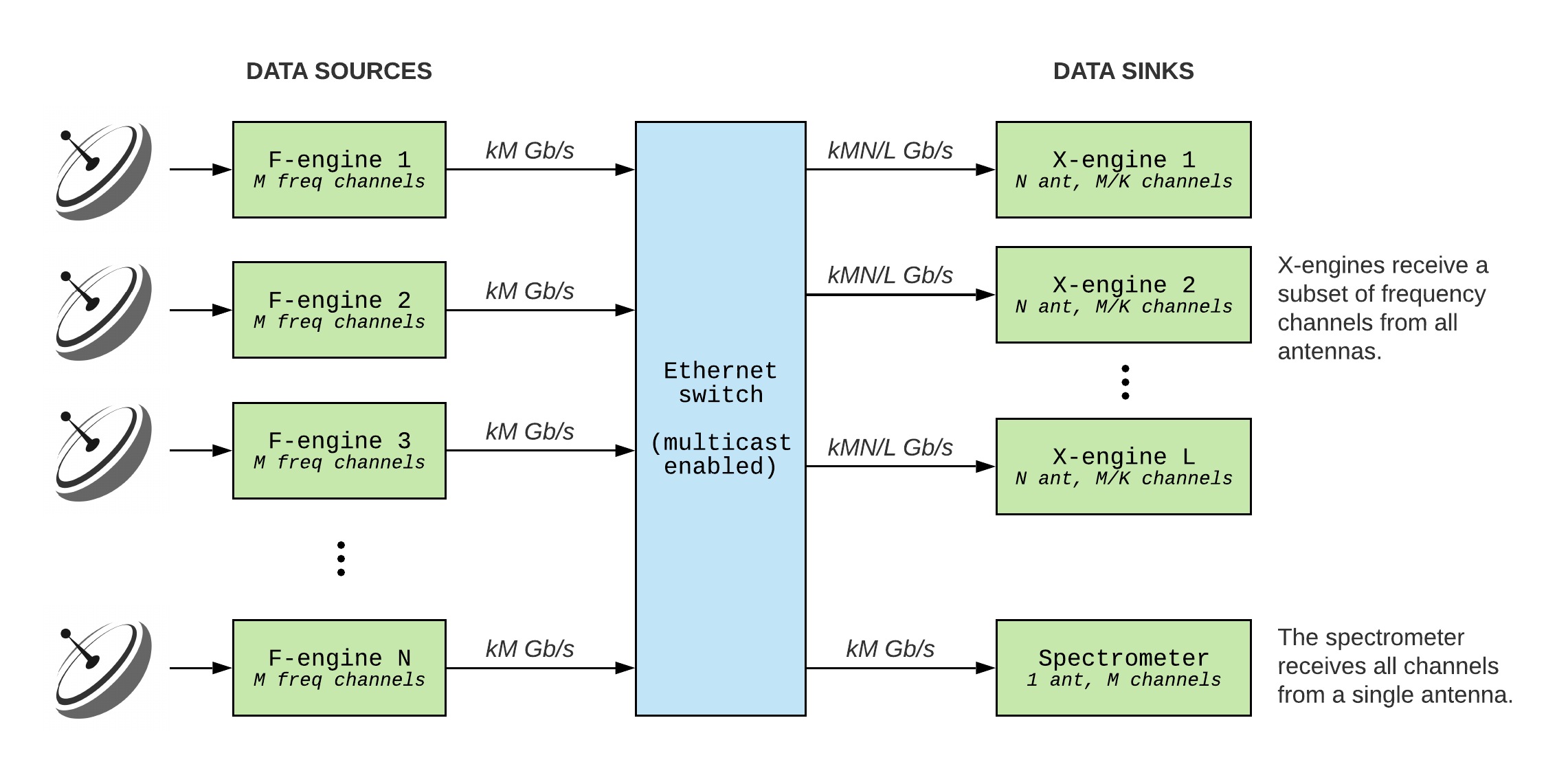}
	\caption{Diagram showing a common approach to building an FX correlator. The outputs of $N$ antennas are digitized and channelized into $M$ channels using a DSP frontend (e.g. FPGA boards). These channels are split into $L$ sub-bands, which are sent over Ethernet to and cross-correlated by $L$ backends running in parallel (e.g. GPU servers). Diagram modified from Fig. 1 of \citet{CASPER:2016}. \label{fig:casper-arch}}
	\end{center}	
\end{figure}

A diagram showing a heterogeneous approach to implementing an FX correlator---where signals are channelized before cross correlation---is shown in Fig.\,\ref{fig:casper-arch}. In this example, $N$ processing nodes are used to channelize data from $N$ antennas. Their output data streams are sent via an Ethernet switch to $L$ correlator nodes running in parallel. In this example the problem of data redistribution (Section\,\ref{sec:transport}) is solved by use of a commercial off-the-shelf network switch. This approach is detailed further in \citet{CASPER:2008} and \citet{CASPER:2016}.

While DSP nodes may run independently from one another, for radio astronomy it is vital that inputs are kept synchronized or else phase coherence will be lost. A common approach is to distribute a single sampling clock to the ADCs on each frontend node. A pulse-per-second, derived from global positioning system (GPS) may also be distributed to the frontend boards, to demarcate an accurate absolute start time. As long as data frames are accurately time tagged, data processing after the ADC may be asynchronous, although frames may need to be buffered in memory as they propagate through asynchronous parts of a pipeline.

\subsection{Common architectures}\label{sec:arch}

The world's radio telescopes are powered by four dominant platforms for digital signal processing: Central Processing Units (CPUs), Graphics Processing Units (GPUs), Field-Programmable Gate Arrays (FPGAs), and Application-Specific Integrated Circuits (ASICs). As their strengths and weaknesses are key to choosing which platform to use, we provide a brief overview here.

\paragraph{ASICs}

\begin{figure}
	\begin{center}
	\includegraphics[width=1.0\columnwidth]{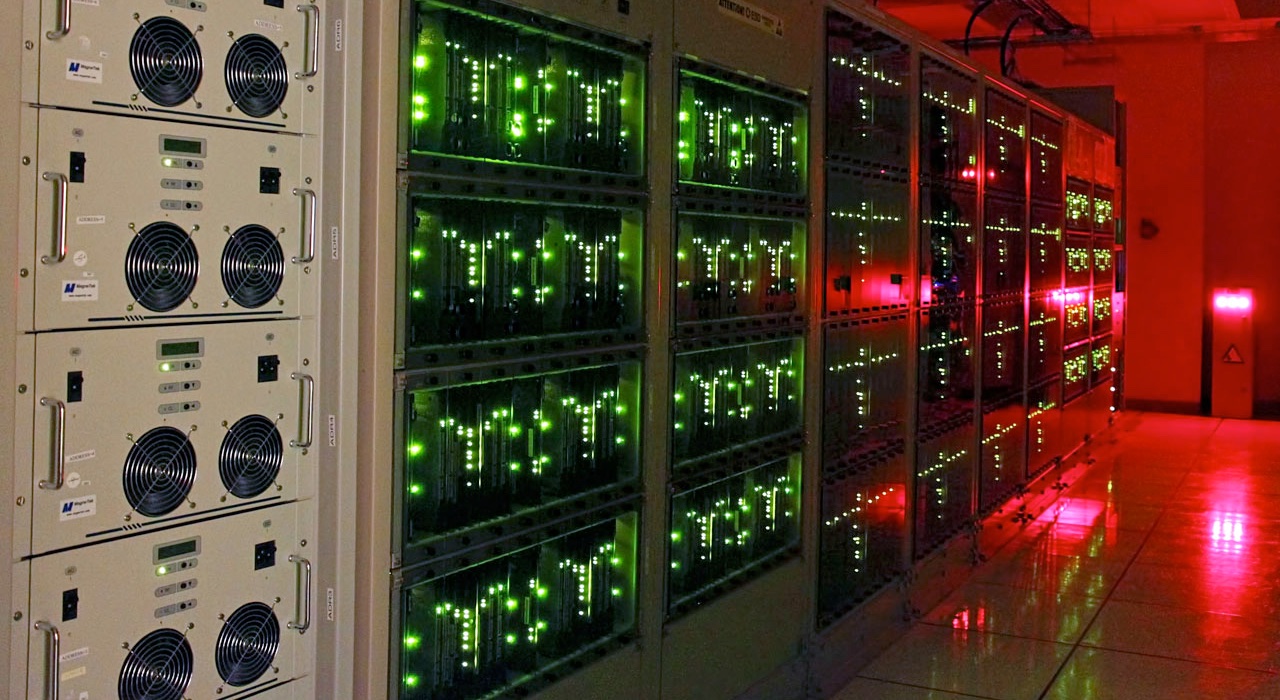}
	\caption{One quadrant of the ASIC-powered ALMA correlator. The ALMA correlator processes up to 16\,GHz bandwidth from 64 antennas. Image credit: ALMA (ESO/NAOJ/NRAO), S. Argandoña\label{fig:alma-correlator}}
	\end{center}	
\end{figure}

As the name suggests, ASICs are digital circuits designed for a specific purpose. ASICs were once common within radio astronomy, and power the  Karl Jansky Very Large Array (JVLA) and Atacama Large Millimeter Array (ALMA, Fig.\,\ref{fig:alma-correlator}) correlators \citep{VLA:2009, ALMA:2011}. However, the one-time cost to fabricate a custom chip on nanometer-scale processes is significant, and economies of scale can be hard to realize for most experiments. For the most part, ASICs have been replaced with off-the-shelf FPGAs and GPUs, which offer a faster development cycle thanks to repurposable codebases and shared toolflows. Indeed, FPGAs are often used for prototyping ASIC designs. Nevertheless, for large-scale projects such as the SKA, ASICs offer significant power savings that could reduce operational costs \citep{Daddario:2016}. The functionality of ASICs is generally defined using a hardware description language (HDL).

\paragraph{CPUs}
The ubiquitous CPU is the most flexible of the four DSP platforms, as it is designed specifically for general-purpose computation. 
The two dominant suppliers of CPUs for server-class computers are Intel and Advanced Micro Devices (AMD). Both Intel and AMD chips are based on the x86-64 instruction set architecture, so codes are intercompatible between the two. For high-performance computing, the Intel Xeon and AMD Ryzen Pro branded processors are marketed, which  support larger amounts of RAM and have advanced features not found on desktop-class chips. An interesting alternative is the IBM POWER9 CPU, which is notable in its support for NVLINK, an 80\,GB/s bus designed for faster memory copy to NVIDIA GPU cards.
 
In mobile and low-power embedded platforms, ARM (previously Advanced RISC Machine) has emerged as the dominant architecture. From a radio astronomy standpoint, ARM processors are commonly found as systems-on-chip (SoC) that directly interface with FPGAs, but they are generally not used for heavy DSP work.

\paragraph{GPUs}

A recent trend is for CPUs to be joined by GPUs as `coprocessors', where computations are offloaded to the GPU. The types of computation required for realistic, high frame-rate graphics are a good match for many astronomy algorithms, where a single instruction (e.g. a multiplication operation) can be applied to multiple data in parallel\footnote{The term single instruction, multiple data (SIMD) was popularized by \citet{Flynn:1972} as part of his seminal taxonomy of computer architectures.}. GPU processors are---to date---not used without a host system.

The use of GPUs for non-graphics processing is known as general-purpose GPU programming, or GP-GPU. As with CPUs, there are two dominant players: NVIDIA and AMD. NVIDIA GPUs are programmed using the Compute-Unified Device Architecture, or CUDA\textregistered\, programming model, whereas AMD requires the use of Open Computing Language (OpenCL), an open-source standard maintained by the Khronos Group. While OpenCL is also supported by NVIDIA GPUs, achieving maximum performance on either vendor's offerings requires targeting of specific instructions and tuning of kernels to match the underlying architectures. For example, the CHIME correlator leverages a fused multiply-add instruction `\texttt{mad24}' only available on AMD GPUs \citep{ChimeXcorr:2015}, whereas the \textsc{xGPU} correlator code leverages a dot-product accumulate `\texttt{dp4a}' instruction only available on some NVIDIA GPUs \citep{xgpu:2013}. As such, writing highly performant code compatible with both vendors is burdensome, and generally code is written in CUDA\textregistered\, if targeting NVIDIA GPUs, or OpenCL if targeting AMD GPUs.

\paragraph{FPGAs}

FPGAs are user-reconfigurable integrated circuits, consisting of an array of programmable logic blocks. These blocks are connected together by a reconfigurable mesh, such that different functionality is achieved by connecting different components together to form a digital circuit. The two dominant FPGA vendors are Xilinx Inc. and Intel\textregistered\, (formerly Altera). As with GPUs, differences in implementation of interfaces and logic elements mean that firmware is designed to target a specific chip. While FPGAs are also used as coprocessors alongside CPUs in high performance computing, for radio astronomy applications they are more commonly found on discrete boards with peripheral interfaces such as ADCs and high-speed Ethernet, to perform first-stage DSP tasks such as channelization and packetization (see Chapter 5). 

FPGAs are well-suited to interface with ADCs and other peripherals due to  high-speed ($>$1Gb/s) transceivers, over which a SerDes (Serializer/Deserializer) link can be setup. To route data from an ADC to an FPGA, the parts are placed onto a circuit board and the relevant pins of the  ADC are connected to the relevant pins of the FPGA; alternatively, the FPGA pins are connected to a mezzanine connector that allows a variety of daughter boards to be connected. Similarly, the FPGA transceivers are used to drive high-speed Ethernet interfaces, over which signals can be transported in Ethernet packets.

FPGAs are most commonly programmed using a HDL, such as VHDL or Verilog, but also support OpenCL. Higher-level tools, such as MATLAB\textregistered\, Simulink and National Instruments LabView, are also available. The Collaboration for Astronomy Signal Processing and Electronics Research (CASPER) have provided the radio astronomy community with open-source FPGA-based hardware and development libraries tools for over a decade \citep{CASPER:2016}, promoting design reuse and lowering entry barriers. Several other FPGA boards for radio astronomy exist, including the ASTRON Uniboard \citep{Uniboard:2010}, the CSIRO Redback and Gemini boards \citep{Redback:2014, Gemini:2017}, and the ICE board from McGill Cosmology Instrumentation Laboratory \citep{ChimeICE:2016}.

\section{Ethernet interconnect}\label{sec:ethernet}

Ethernet is an industry-standard collection of computer networking technologies used in virtually all computer networks. Computer networking is a vast topic, well outside the scope of this chapter; nevertheless a general understanding is important to fully appreciate implementation concerns of Ethernet-based data redistribution, which has become commonplace. To that end, a gentle introduction is provided here to guide our discussion. We refer the reader to \citet{TCPIP:1994} and \citet{TCPIP:2005} for a comprehensive overview; the latter is now freely available online at \url{http://www.tcpipguide.com}. Our goal for this section to is to give the reader an idea of the steps required to design a well-scoped and fully-functional data redistribution network for a small-to-medium size DSP system. 

\subsection{High-speed Ethernet technologies}

Ethernet interconnect uses either passive copper links, or fiber optic links driven by active transceivers. For short distances ($<$10 m), more economical copper cables can be used, but longer distances require fiber optic links, which can span several kilometres.  

Ethernet defines a variety of connector types and speeds. As of writing, 10\,GbE and 40\,GbE are widespread and economical, with 100\,GbE becoming increasingly available. The SFP+ (small form-factor pluggable transceiver) is most widespread for 10 GbE, whereas and QSFP+ (quad SFP) is used for 40\,GbE. A 40 GbE link can be split into 4$\times$10\,GbE links, meaning a 32-port 40\,GbE switch can operate as a 128-port 10\,GbE switch---if supported by the switch firmware.

Two common multi-interface devices are used in Ethernet networks: switches and routers. In general, only switches are required for data redistribution networks. Switches work by identifying what devices are connected, then directing incoming packets to their destination via the most suitable port. Routers act more like a gateway between two networks, sending packets from one network to another.

\subsection{TCP/IP}\label{sec:tcpip}

TCP/IP refers to a family of transport protocols, including TCP (Transport Control Protocol), UDP (User Datagram Protocol), and IP (Internet Protocol). UDP and TCP are the prevailing protocols used to transport data across a network, and sit on top of IP. IP defines how data are relayed across a network: data are packaged up as discrete `datagrams' that can be lost, duplicated or delivered out-of-order. This non-guaranteed strategy contrasts starkly with simpler electrical signalling schemes, as would be used in backplanes to connect transceivers of neighboring boards. It introduces an extra level of complexity and processing overhead, but offers advantages for large networks, as it offers a means to dynamically share a link with multiple hosts.

Within TCP/IP, all network interfaces are assigned an IP address---a numerical label akin to a telephone number---by the network administrator. Every interface also has to have a MAC address, a 48-bit identifier that is generally set by the manufacturer so as to be globally unique. MAC addresses are usually expressed as six hexadecimal pairs, e.g. \texttt{00:00:00:00:00:00}. IP (v4) addresses are 32 bits\footnote{The latest version IPv6, uses 128-bit addresses expressed with a hexidecimal notation, e.g. \texttt{2001:0db8:0000:0000:0000:8a2e:0370:7334}; this length is necessary to service the global internet, but not for private internal networks where the shorter IPv4 addresses may still be used.}, expressed in a dot decimal notation of four integers between 0-255, separated by dots, e.g. \texttt{192.168.0.1}. In a private network, you can decide yourself how to allocate IP addresses, but interfaces connected to the broader internet must be assigned IP addresses by the internet provider.

Along with the IP address, every interface must have a subnet mask defined. A subnet mask is used to subdivide the 32-bit IP address range into sub-networks in which traffic is contained; see Chapter 9 of \citet{TCPIP:2005}.

\paragraph{An example IP address table}\label{sec:redist-impl}

\begin{table}[]
\begin{center}
\caption{Example IP allocation for a small correlator.  \label{tab:ipalloc}}
\begin{tabular}{llll}
Device     & IP Address    & Subnet mask   & MAC Address       \\ \hline
F-engine 1 & \texttt{192.168.10.101} & \texttt{255.255.255.0} & \texttt{02:00:00:00:00:01} \\
F-engine 2 & \texttt{192.168.10.102} & \texttt{255.255.255.0} & \texttt{02:00:00:00:00:02} \\
F-engine 3 & \texttt{192.168.10.103} & \texttt{255.255.255.0} & \texttt{02:00:00:00:00:03} \\ 
F-engine 4 & \texttt{192.168.10.104} & \texttt{255.255.255.0} & \texttt{02:00:00:00:00:04} \\
X-engine 1 & \texttt{192.168.10.201} & \texttt{255.255.255.0} & \texttt{02:01:00:00:00:01} \\
X-engine 2 & \texttt{192.168.10.202} & \texttt{255.255.255.0} & \texttt{02:01:00:00:00:02} \\ 
X-engine 3 & \texttt{192.168.10.203} & \texttt{255.255.255.0} & \texttt{02:01:00:00:00:03} \\ \hline
\end{tabular}

\end{center}
\end{table}

One of the first tasks to bringing up a Ethernet-based data redistribution network is to allocate IP addresses to all network interfaces. To put IP addressing into context, consider an example FX correlator, with four F-engines running on FPGAs, connected to three X-engines servers with GPUs. Note that we are following the approach shown in Fig.\,\ref{fig:casper-arch}, shown as a stream processing diagram in Fig.\,\ref{fig:cornerturn}. 

An example IP allocation table for the putative correlator is shown in Tab.\,\ref{tab:ipalloc}. We have chosen to use the \texttt{192.168.10.0} network, and set a subnet mask \texttt{255.255.255.0}.

\subsection{TCP and UDP}

As mentioned above, TCP and UDP are the prevailing protocols used for network data communications. Both protocols send data as a series of datagrams, or `packets' with roughly $\sim$kB sizes. The main difference between the two is that TCP checks if data are transferred successfully, and if not, will resend missing data. For large, complex networks like the internet, this makes TCP a reliable method to transfer data; however, it has significant overhead due the requirement that the receiver must send acknowledgements and requests to resend missing data back to the transmitter. Data must also be buffered so that it can be retransmitted if required. In contrast, UDP is an unguaranteed, one-directional stream of packets that the receiver does not need to acknowledge with a response. 

UDP is therefore easier to implement, and faster overall transmission speeds can be achieved due the the lower overhead. While packet loss during transmission is a potential issue, this is rarely encountered in well-scoped data redistribution networks: packet loss is a symptom of long (i.e. high latency) links with contention due to congestion, and/or hardware failure.

\subsection{UDP Datagram Structure}

\begin{figure}
	\begin{center}
	\includegraphics[width=1.0\columnwidth]{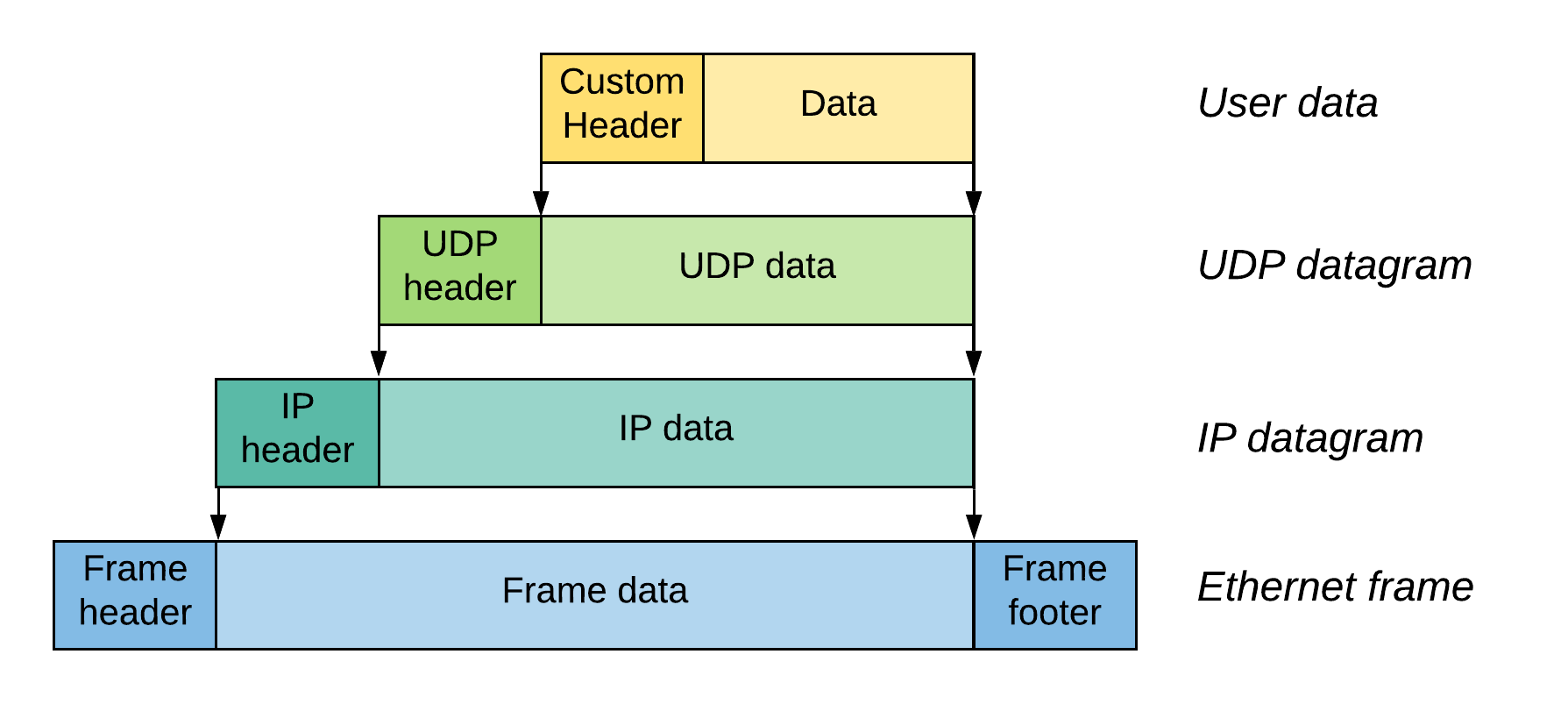}
	\caption{The layers within an Ethernet frame for a UDP datagram. Data units are nested within the data payload of each lower layer. Within the UDP data payload, a user may define their own data structure, such as a header plus data payload. \label{fig:udp-stack}}
	\end{center}	
\end{figure}

Units of data are passed over UDP as packets, or more correctly \emph{UDP datagrams}\footnote{The term `packet' is used generically to refer to any type of IP message. TCP calls its messages segments, while UDP calls its messages datagrams.}. UDP datagrams are themselves embedded into \emph{IP datagrams}, which are themselves embedded into an \emph{Ethernet frame} (not to be confused with the use of frame elsewhere in this chapter), rather like nested matryoshka dolls (see Fig\,\ref{fig:udp-stack}). 

At each layer, there is a header and a data payload. Headers are several bytes in size (IPv4 header is 20 bytes, and UDP header is 8 bytes) while payloads can be thousands of bytes---the maximum size of an Ethernet frame is 1500\,B, or up to 9000\,B if `jumbo frames' are enabled in the Ethernet hardware. Due to the headers, small datagrams have appreciable overhead, so larger datagrams are preferred for high throughput applications. In addition, the number of packets per second decreases as packets get larger, which is beneficial when capturing large data volumes (see \S\,\ref{sec:packet-capture}).

While this is admittedly complex, as long as we are happy to adhere to the TCP/IP standard, methods for handling the lower levels come `baked in' to all Ethernet devices. On FPGAs, Ethernet IP cores are provided by vendors and third parties. As such, the astronomer/engineer only needs to focus on how to structure their data within the UDP datagram's payload: the yellow part of Fig.\,\ref{fig:udp-stack}.

A common choice is to set aside a few bytes as the start of the UDP data payload and define a custom header that describes the data well enough to reconstruct the data stream. At its simplest, the header would just be a counter so that packets can be reordered correctly in case they are received out-of-order. If there are multiple nodes transmitting data, a few bytes of header could be allocated for a unique board identifier (e.g. an 8-bit number between 0-255). If a single frame from the data stream cannot fit within the maximum packet size of 8972\,B---for example, if there are 32,768 channels of 32-bit datatype---then extra information is needed to identify what chunk of the frame is being transmitted, so the frame can be faithfully reconstructed. While specifications for streaming protocols over UDP exist, such as SPEAD\footnote{Streaming Protocol for the Exchange of Astronomical Data, \url{https://spead2.readthedocs.io/}}, it is common for the UDP payload data structure to be custom to a given experiment.

\subsection{Multicast}

IP Multicast is a useful technique that allows incoming data sources to be distributed to multiple data sinks. Instead of sending to the IP address of a single node, data are sent to a multicast IP address (between {\tt 224.0.0.0} to {\tt 239.255.255.255}). Any nodes that want to receive these data `subscribe' to the multicast data by joining the multicast group (defined by the multicast IP address chosen). For this to work, the network must have a device acting as a multicast router, which uses Internet Group Management Protocol (IGMP) to establish group membership. Network switches use IGMP snooping to maintain a list of which links it should re-transmit multicast packets over (alternatively, switches can re-broadcast multicast packets over all ports, which is generally undesirable). 

Multicast is used in the MeerKAT telescope as a method to allow user-supplied equipment access to data streams from the first-stage FPGA boards. They also use multicast as part of their corner-turn operation: subbands are sent to different multicast groups.

\section{First-stage data processing}\label{sec:frontend}

\begin{table}[]
\caption{Aggregate data rates after digitization for selected radio telescopes.   \label{tab:data-rates}}
\footnotesize
\begin{tabular}{lcccccccc}
\hline
Telescope & $F_s$  & $N_{\rm{bits}}$ & $N_{\rm{ant}}$ & $N_{\rm{rxb}}$ & $N_{\rm{pol}}$ & $N_{\rm{sub}}$ & DR     & Reference \\
          & (MHz) &         &        &        &        &        & (Tb/s) &  \\
          \hline
          \hline
ASKAP     & 608   & 12      & 36     & 188    & 2      & 1      & 98.8   &  A \\
CHIME     & 800   & 8       & 1024   & 1      & 2      & 1      & 13.1   &  B \\
ALMA      & 4000  & 3       & 64     & 1      & 2      & 4      & 6.1    &  C \\
MeerKAT   & 1712  & 10      & 64     & 1      & 2      & 1      & 2.2    &  D \\    
SMA       & 4000  & 8       & 8      & 1      & 1      & 4      & 2.0    &  E \\
JVLA      & 2048  & 8       & 26     & 1      & 2      & 2      & 1.7    &  F \\
\hline
\multicolumn{8}{l}{$^{\rm{A}}$ \citet{Askap:2012} }\\           
\multicolumn{8}{l}{$^{\rm{B}}$ \citet{ChimeFRB:2018,ChimeICE:2016} }\\           
\multicolumn{8}{l}{$^{\rm{C}}$ \citet{ALMA:2011} }\\     
\multicolumn{8}{l}{$^{\rm{D}}$ \citet{MEERKAT:2009} }\\     
\multicolumn{8}{l}{$^{\rm{E}}$ \citet{SMA:2016} }\\     
\multicolumn{8}{l}{$^{\rm{F}}$ \citet{VLA:2011} }\\     
\end{tabular}
\end{table}

Now that we have introduced common technologies, we now discuss how these technologies may be used to implement stream processing systems. The first step in a radio astronomy processing pipeline is to convert incoming signals from analog to digital, using an analog-to-digital converter (ADC). This step is detailed in Chapters 3 and 5, along with common first-stage data processing approaches; here, we provide a short overview to highlight the challenges faced downstream. 

\subsection{Data rates}

ADCs act as a data sources in a stream processing pipeline. The data rate from the ADC depends upon the sampling rate, $F_s$, and the number of bits per sample, $N_{\rm{bits}}$:
\begin{equation}
	{{\rm{Data\,rate\,(b/s)}}} = F_s \times N_{\rm{{bits}}}
\end{equation}
For example, a 1\,Gsample/s ADC with 8-bit resolution outputs an 8\,Gb/s data stream. The Nyquist–Shannon sampling theorem dictates that the ADC sampling rate must be twice the analog bandwidth $B$ of the incoming signal, $F_s = 2\times B$. For current-generation radio telescopes, receiver bandwidths may span several gigahertz.

In order to capture the two orthogonal polarization states ($N_{\rm{pol}}=2$), two ADCs are required per telescope receiver. Some telescopes, such as ASKAP, have multiple receivers per antenna: the ASKAP phased array feed consists of hundreds of receivers that are digitally combined to form beams on the sky \citep{Askap:2012}. Similarly, receivers may split their bandwidth over several sub-bands. The aggregate ingest data rate for an array of radio telescopes is thus given by
\begin{equation}
	{{\rm{Data\,rate\,(b/s)}}} = (F_s \times N_{\rm{{bits}}}) \times N_{\rm{ant}} \times N_{\rm{pol}} \times N_{\rm{rxb}} \times N_{\rm{sub}}, 
\end{equation}
where $N_{\rm{rxb}}$ is the number of receiver elements per antenna, and $N_{\rm{sub}}$ is the number of sub-bands. 

Table\,\ref{tab:data-rates} shows aggregate data rates for a selection of radio telescopes where total ingest exceeds 1\,Tb/s. These rates may be exceeded at parts in the pipeline, due to increased datatype size (e.g. bit growth within a FFT), or an increase in frame size (e.g. after cross-correlation across antenna pairs). Nevertheless, data rates must eventually be reduced by many orders of magnitude in order to form science data products that can be written to disk.

\subsection{Channelization}

After digitization, it is common for input signals to be split into channels using a FFT or polyphase filterbank (PFB; see \citet{Price:2016}, \citet{TMS:2017} \S8.8, for an overview). Channelization is motivated by two reasons. Firstly, many processing tasks, including cross-correlation and beamforming, can be parallelized across frequency channel. Secondly, as shown in Figure\,\ref{fig:casper-arch}, a data frame can be split across channels and distributed across multiple processing boards. 

The channelizer part of a stream processing pipeline is commonly referred to as an `F-engine'. This term is used broadly to include gain correction, requantization of signals to lower bitdepth, and preparation of the data stream for signal transport. F-engines are explored in more detail in Chapter\,5.

\section{Data redistribution}\label{sec:transport}

Getting data from multiple source nodes and routing it to multiple sinks nodes is a non-trivial task for large installations. In this section, we discuss common approaches to data redistribution, with a focus on the use of Ethernet for multi-node stream processing.

\subsection{The corner-turn problem}

\begin{figure}
	\begin{center}
	\includegraphics[width=1.0\columnwidth]{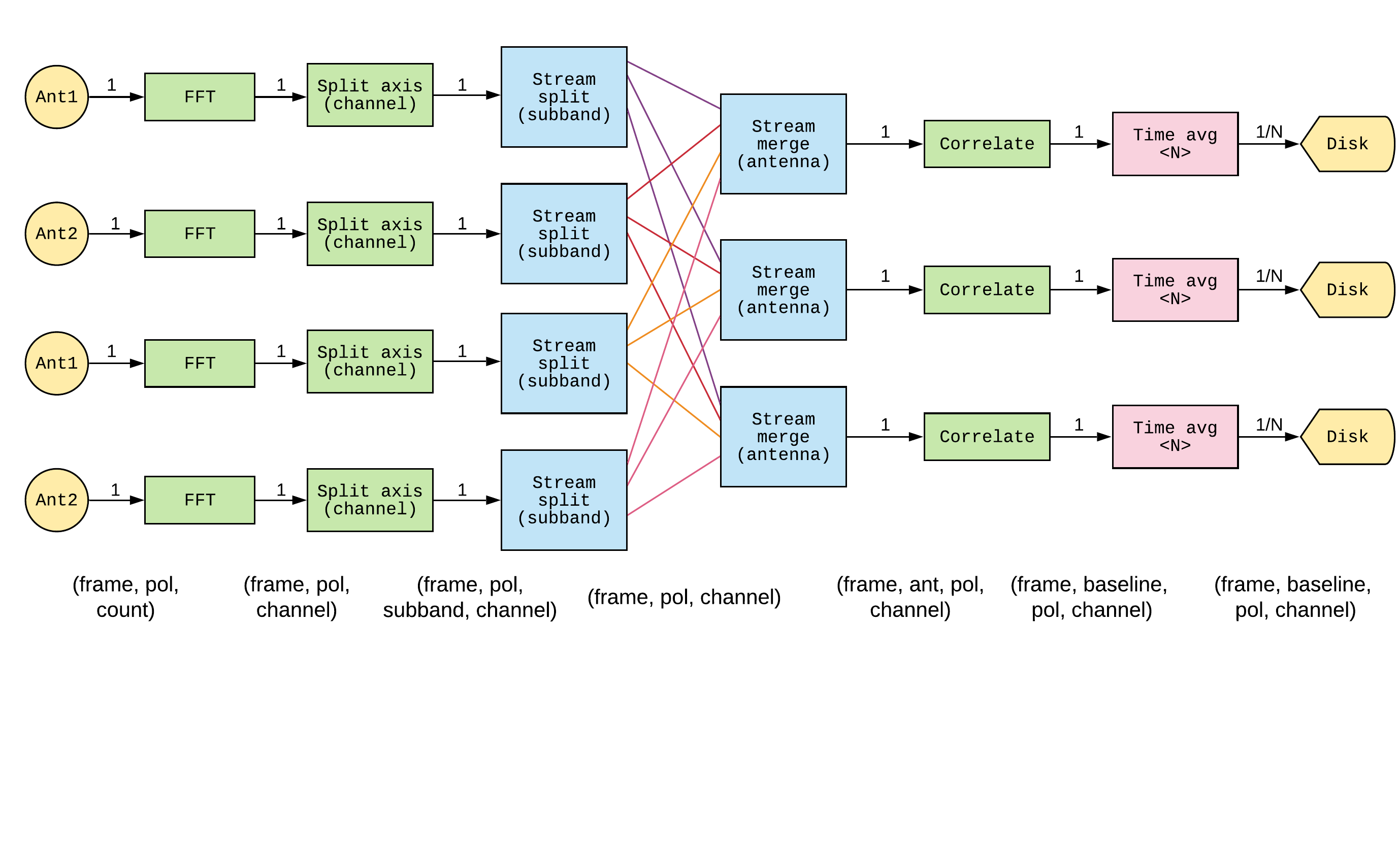}
	\caption{A stream processing pipeline showing a four-element cross correlator with a corner-turn operation. Here, after the FFT operation the frame is split into three subbands. Three separate correlator streams run in parallel, each processing one subband. This stream splitting then merging is used to group a fraction channels from every antenna input, and is known as a corner turn; as the number of antennas increases the corner-turn operation can become difficult. \label{fig:cornerturn}}
	\end{center}	
\end{figure}

 The problem of data redistribution is exemplified by correlators with a large number antenna inputs. In a so-called FX correlator, the data stream from each antenna is channelized into $M$ channels by a filterbank, referred to as an `F-engine'. The F-engine is fed into an `X-engine' that correlates the signals from all antenna pairs together. In the F-engine, each antenna is independent, so a separate F-engine can be run on each of $N$ antenna inputs, to produce $M$ channels. The X-engine computes the cross-correlation for every pair of antennas, but each channel can be processed independently. The X-engine can therefore be split into $L$ separate X-engines, each processing $M/L$ channels. 

An stream processing diagram for a simple FX correlator is given in Fig.\,\ref{fig:cornerturn}, with $N$=4 F-engines connected to $L$=3 X-engines. The data streams are split into subbands and then merged by antenna; this process is known as a \emph{cornerturn}, and is conceptually similar to a $N\times L$ matrix transpose. 

As $N$ and $L$ increase, the corner-turn becomes increasingly difficult, by virtue of the $N\times L$ stream split/merge operations required. Adding to the complexity are the massive volumes of data that are being transported, and limitations on the data rate each link can handle. Choosing an appropriate and scalable data redistribution scheme is an important design choice in any radio astronomy instrument.

\subsection{Data backplanes}

One method to redistribute data is to use a custom backplane into which processing boards are plugged. The backplane implements a network of links, generally using differential signalling over copper lines. This method is only appropriate for FPGAs and ASICs, as their transceivers can drive the differential lines of the backplane to be routed to  other FPGAs or ASICs within the system. Backplanes also provide a convenient way to route power and clock signals to the boards. Nevertheless, there are limitations to how large the redistribution network can grow, as the number of transceivers on any FPGA is finite; plus, there are physical routing considerations as more boards are added to the backplane. Only devices specifically designed for the backplane may be connected, which limits their portability and compatibility with other technologies.

\paragraph{The CHIME backplane.} An example of a modern backplane is that used in the CHIME telescope \citep{ChimeICE:2016,ChimeXTurn:2016,ChimeFRB:2018}. CHIME is the largest radio interferometer build to date, based on the metric of number of inputs squared times bandwidth ($N^2B$), processing data from $N$=2048 inputs, each with $B$=400\,MHz. CHIME uses the `ICE' FPGA-based signal processing and networking system, in which up to sixteen FPGA motherboards are plugged into 9U ($\sim$40\,cm) rack-mountable `crates'. Each crate has a custom backplane that provides high-speed, full-mesh connectivity between the sixteen boards; 64x 10 Gb/s Ethernet links are also provided to interconnect crates (4 links per FPGA board). The backplane is constructed from 24 substrate layers, and interfaces the FPGA boards using high-density Impact-series Molex connectors that can pass up to 25 Gb/s. 

A total of 6.6 Tb/s of data flows through the CHIME corner-turn system before being fed into 256 GPU-based correlator nodes. The corner-turn is done in several stages. The first is conducted on the FPGA boards, each of which processes 16 antenna streams. The second stage uses the backplane full-mesh network, to gather subbands from 256 antenna inputs per crate. Thirdly, multiple crates are connected using the 10 Gb/s Ethernet links, and each FPGA board ends up with 32 channels from 512 inputs; two more corner-turn stages are then conducted on the GPU servers. 

\subsection{Packetized Ethernet interconnect}

An alternative method is to connect all boards via a high-speed Ethernet network. This approach is becoming increasingly common, as GPU-based instruments gain traction and Ethernet switches and interface cards become cheaper and faster. 

The simplest method is to connect all interfaces, i.e. all data sources and all data sinks, to a single switch. Some care is needed to ensure that instantaneous packet rates are not too `bursty', so as to momentarily exceed the output port's data rate; many switches have hardware flow control to alleviate this issue, but flow control buffers on ports are often small.

\paragraph{The MeerKAT data network.} The MeerKAT telescope \citep{MEERKAT:2009} implements an impressive data redistribution network \citep{MEERKAT:2018} that allows multiple racks of second-stage `user supplied equipment' (USE) to access the incoming data streams from 64 antennas. USE subscribes to incoming data streams using multicast join requests; data from the first-stage signal processors is divided into subbands and each subband is sent to a separate multicast IP. 

\begin{figure}
	\begin{center}
	\includegraphics[width=1.0\columnwidth]{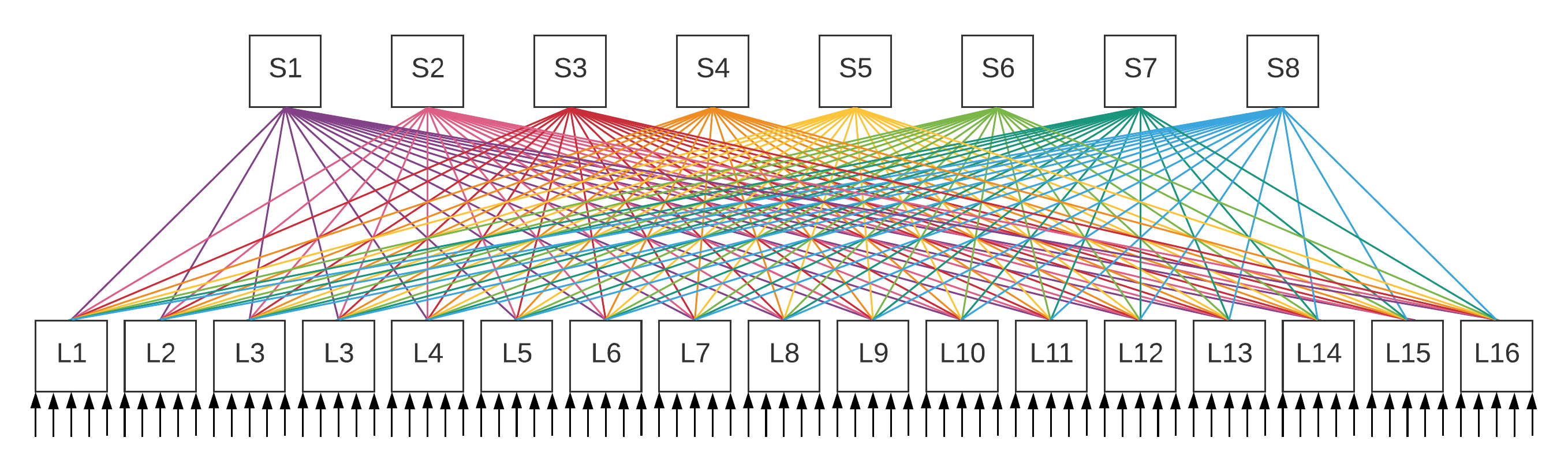}
	\caption{A network with a folded Clos topology, with eight spine nodes connected to sixteen leaf nodes. Equipment only connects to leaf switches (shown with black arrows). \label{fig:clos-meerkat}}
	\end{center}	
\end{figure}

Essentially, the MeerKAT data redistribution acts as if one giant switch, with 648 available ports. This virtual switch is implemented using a folded Clos topology (see example in Fig.\,\ref{fig:clos-meerkat}), where there are 36 `leaf' switches connected to 18 `spine' switches, all switches are the same hardware. Each leaf switch is connected to every spine switch, but not to other leaf switches. Each USE rack may have a leaf switch physically located within it, for convenience.  

As there are multiple paths a packet could take, a packet routing scheme must be implemented to avoid loops and saturation. MeerKAT uses the Open Shortest Path First (OSPF) IP protocol, but manual routing tables can also be implemented.

\section{Second-stage processing}\label{sec:backend}

Given the widespread use of CPU/GPU servers for second-stage processing, in this section we discuss factors relevant to achieving high performance on CPU/GPU servers, and detail common approaches. 

\subsection{Performance modelling} 

To keep hardware costs down, each node within a second-stage processor should capture and process as large a bandwidth as possible. The limit will depend on whether the processing pipeline is compute bound or bandwidth (I/O) bound. In a compute-bound task, the amount of data that can be processed is limited primarily by the time taken for computational operations to be performed. Conversely, in a I/O-bound task, the movement of data from source to sink is the primary performance limitation. Both bounds are encountered in radio astronomy pipelines, depending on the telescope specifications and signal processing requirements. 

The roofline performance model \citep{Roofline:2009} offers an intuitive way to determine whether blocks in a pipeline are likely to be compute or I/O bound. For a given block, the achievable number of operations per second (ops/s) is, to first order, given by:
\begin{equation}
{\rm Achievable\,ops/s=\emph{min}\left\{ \begin{array}{c}
{\rm peak\,performance}\\
{\rm peak\,bandwidth\times operational\,intensity}
\end{array}\right.}
\end{equation}
where the underlying hardware specifications set the peak computational performance, and relevant I/O bandwidths (memory, PCI bus, network interfaces, etc) set the peak bandwidth. The term `operational intensity' refers to the number of computational operations required per byte of I/O traffic.

Achieving the peak performance of the compute hardware can be a difficult task. Some useful avenues or attack, starting with efficient packet capture, are outlined below. 

\subsection{Packet capture}\label{sec:packet-capture}

High-speed packet capture is arguably the most difficult part of bringing up a data processing pipeline on a compute server. It invariably requires programming in a low-level language such as C, as scripting languages such as Python are not performant. 

The easiest way to explain why is to take a dive into some basic C code to capture a packet. In Unix/BSD, packets are captured by creating a \emph{socket}, which is a communication endpoint that acts as a file descriptor. Let's take a look at the synopsis from the Linux programmer's manual\footnote{\url{http://man7.org/linux/man-pages/man2/socket.2.html}}:
\begin{minted}{c}
#include <sys/types.h>          /* See NOTES */
#include <sys/socket.h>

int socket(int domain, int type, int protocol);	
\end{minted}
Creating a socket requires three things to be set: a {\tt domain}, a {\tt type}, and a {\tt protocol}. 
The {\tt domain} specifies which communication protocol we wish to use; for TCP/IP we choose {\tt AF\_INET}. For UDP, we set the {\tt type} argument to {\tt SOCK\_DGRAM} (or {\tt SOCK\_STREAM} for TCP), and the {\tt protocol} argument should be set to {\tt 0} (not required for UDP). So a UDP socket is created with the call:
\begin{minted}{c}
int socket(AF_INET, SOCK_DGRAM, 0);	
\end{minted}
Once created, (and skipping several lines of setup), the {\tt recvfrom}\footnote{\url{https://linux.die.net/man/2/recvfrom}} call is used to receive a packet from the socket:
\begin{minted}{c}
ssize_t recvfrom(int sockfd, void *buf, size_t len, int flags,
                 struct sockaddr *src_addr, socklen_t *addrlen);	
\end{minted}
where {\tt sockfd} is a socket you've created, {\tt *buf} is a pointer to a memory buffer into which the packet payload should be copied, {\tt len} is the size of the buffer, {\tt flags} are optional flags, {\tt sockaddr} holds the IP address and port to receive data from, and {\tt addrlen} is just the size of the {\tt sockaddr} struct (i.e. {\tt sizeof(struct sockaddr)}). The {\tt recvfrom} call is handled by the operating system kernel. When the network interface card (NIC) receives a packet, its driver will issue a hardware interrupt request (IRQ) to the kernel, to let it know data has arrived.  

Receiving a stream of UDP data thus requires writing a loop and calling {\tt recvfrom} multiple times, to get the packet from the NIC into kernel space, and then up into user space where your packet capture code runs. Once the packet is in user space, one still needs to parse the information in the UDP packet to make sure they have been received in the correct order, and then place their data in the right spot in memory to recreate the data frames. All this must be done before the next packet is received, or multiple processing threads have to be used to keep up. 

At high packet rates, keeping up can become difficult. For example, consider a 10 Gb/s stream of packets of size 1000\,B. The number of packets received per second is 1.25M, which means 1.25M calls to {\tt recvfrom} are made, each with corresponding interrupts. The sheer number of kernel calls can quickly become a bottleneck, and the system may not keep up with the data rate. 

\subsection{Kernel bypass packet capture}

When performance becomes an issue (likely for $>$10\,Gb/s streams), the first fix one might attempt is increasing the size of the packets from the data source. Moving from a 1000\,B packet to a 8000\,B packet would decrease the packet rate by a factor of 8$\times$. Another approach would be to use what is called a `raw socket'---which bypasses the TCP/UDP layer and gives  access to lower-level, un-extracted packets---with packetmmap\footnote{\url{https://sites.google.com/site/packetmmap/}} to grab multiple packets with one call. We advise against this approach, in favor of the use of accelerator frameworks designed specifically for high-speed packet capture, such as DPDK\footnote{Data Plane Development Kit, \url{https://www.dpdk.org/}}, and ibverbs\footnote{\url{http://rdmaconsortium.org/}}, which allow for data to be copied to memory without any kernel calls, known as \emph{kernel bypass}. Bypassing the kernel to copy data is known as a zero copy, and when performed on data incoming from another device, is known as remote direct memory access (RDMA). Both DPDK and ibverbs require RDMA-capable network cards and drivers; for example, Mellanox NICs require their proprietary VMA\footnote{\url{https://www.mellanox.com/page/software_vma}} library to be installed in order to use ibverbs.

\paragraph{Example implementations.} Table \ref{tab:packet-capture} shows data ingest rates for a selection of second-stage signal processing instruments that capture above 20\,Gb/s per server. Note that computational bounds and technical considerations (e.g. power and cooling) are important considerations, so higher ingest rates do not directly reflect higher computational efficiency. The ibverbs approach is used by packet capture code running for the Parkes Ultra-Wideband Low receiver, to capture 24.8 Gb/s per server \citep{UWL:2019}, and also in the SPEAD2 packet capture library\footnote{https://spead2.readthedocs.io/} used by the MeerKAT telescope. In the TRAPUM backend for transient studies with MeerKAT \citep{Trapum:2016}, up to 56\,Gb/s is captured per server, using a pair of 40\,GbE NICs. The Hydrogen Epoch of Reionization Array \citep[HERA,][]{HERA:2017} uses ibverbs to capture 2$\times$34\,Gb/s per server, again using a pair of 40\,GbE NICs. The CHIME X-engine uses DPDK to capture 25.6 Gb/s per server \citep{ChimeKotekan:2015}. The LEDA correlator \citep{LEDA:2014} ingests 21.4\,Gb/s, without the use of an accelerator library. 

\begin{table}[]
\caption{Data ingest rates (per server) for selected second-stage signal processing instruments.  \label{tab:packet-capture}}
\footnotesize
\begin{tabular}{lccccc}
\hline
Instrument & Ingest rate   & Packet size & Capture method & Data pipeline  & Ref. \\
           & (Gb/s/server) &    (B)      &                &                &        \\
          \hline
          \hline
HERA            & 68.0    &  4608       &   ibverbs        & HASHPIPE      & A \\
TRAPUM          & 56.0    &  1024       &   ibverbs/SPEAD2 & PSRDADA       & B \\
CHIME           & 25.6    &  8592       &   DPDK           & kotekan       & C \\
Parkes UWL      & 24.8    &  8272       &   ibverbs        & PSRDADA       & D \\
LEDA            & 21.4    &  7008       &   socket         & PSRDADA       & E \\
\hline
\multicolumn{6}{l}{$^{\rm{A}}$ \citet{HERA:2017}, J. Hickish (personal comms)  } \\
\multicolumn{6}{l}{$^{\rm{B}}$ \citet{Trapum:2016}, E. Barr (personal comms)  } \\
\multicolumn{6}{l}{$^{\rm{C}}$ \citet{ChimeKotekan:2015, ChimeFRB:2018} }\\
\multicolumn{6}{l}{$^{\rm{D}}$ \citet{UWL:2019}  } \\
\multicolumn{6}{l}{$^{\rm{E}}$ \citet{LEDA:2015}  } \\

\end{tabular}
\end{table}

\subsection{Ring buffers}

Once captured from the NIC, data is captured into a buffer, from which a processing pipeline can grab data to process. 

A \emph{ring buffer}, or circular buffer, is a data structure used to emulate a wrap-around in memory. Ring buffers are contiguous blocks of memory that act as if they are circular, with no beginning or end, instead of 1-dimensional. In an asynchronous pipeline, when a write process and read process are sharing a contiguous block of regular memory, the writing process has to either stop when it reaches the end of the block, allocate more memory, or go back to the start address. In a ring buffer, a write process will automatically wrap around to the start address block of memory and will continues to write seamlessly. 

A ring buffer will generally be an integer multiple of the size of the data frames being buffered, plus a `ghost region' to help synchronize the beginning and end of the buffer. Ring buffers can be used to ensure that a memory address being read by a process is not overwritten during the read by the writing process. However, if the read process is too slow, the writer will need to chose whether to stop writing until the reader catches up, or to continue writing and risk skipping data frames.  

Ring buffers are as a standard data structure for buffer management in stream processing frameworks. An alternative approach, ping-pong buffering, can be considered a specific implementation of a ring buffer with exactly two fixed-length elements. In general, ring buffers are not required between processing elements in a FPGA-based pipeline, as operations are guaranteed to occur synchronously. 

\subsection{CPU/GPU pipeline frameworks}

A number of open-source CPU/GPU pipeline frameworks have been deployed in radio astronomy systems, including:

\begin{itemize}
	\item PSRDADA\footnote{\url{http://psrdada.sourceforge.net/}}, where DADA stands for Distributed Acquisition and Data Analysis, was designed for recording and stream processing pulsar data. PSRDADA implements ring buffers in shared memory, and provides a C API for reading and writing from the buffers, along with monitor and control scripts. Pipelines are created by spawning multiple processes that communicate through ring buffers. The Swinburne Pulsar Instrumentation Package, SPIP\footnote{\url{https://github.com/ajameson/spip}} extends PSRDADA an object-oriented approach. PSRDADA is used  in \citep{UTMOST:2017},
	\item HASHPIPE\footnote{\url{https://casper.ssl.berkeley.edu/wiki/HASHPIPE}}, the High Availibility Shared Pipeline Engine, provides a C API for designing pipelines, where processing blocks are run in separate threads. As with PSRDADA, processing blocks are joined with ring buffers. HASHPIPE a derivative of an earlier pipeline called GUPPI, the Green Bank Ultimate Pulsar Processing Instrument \citep{GUPPI:2009}.    
	\item \textsc{Kotekan}\footnote{\url{http://lwlab.dunlap.utoronto.ca/kotekan/}} \citep{ChimeKotekan:2015} is a C/C++ pipeline used in the CHIME telescope, which launches processing blocks in threads. Ring buffers are implemented for both CPU and AMD GPUs. 
	\item \textsc{Bifrost}\footnote{\url{http://ledatelescope.github.io/bifrost/}} \citep{bifrost:2017} is written in C++ and Python, where Python is used as a high-level wrapper to ring buffers and kernels implemented using C++ and CUDA. \textsc{Bifrost} comes with a collection of configurable GPU-accelerated processing blocks, that are combined into a performant pipeline using the high-level Python interface. At runtime, ring buffers are automatically created, and blocks are connected through these ring buffers as a directed graph. Bifrost is used to process data from the 256-antenna Long Wavelength Array station at Sevilleta (LWA-SV), which implements a beamformer, correlator and direct imaging correlator \citep{EPIC:2017}. 
	
\end{itemize}
Other efforts include PELICAN\footnote{\url{https://github.com/pelican/pelican}} \citep{pelican:2015}, which is designed for static, quasi-realtime purposes. Outside radio astronomy, the gstreamer\footnote{\url{https://gstreamer.freedesktop.org/}} framework has been used in gravitational wave detection \citep{GravStreamer:2017}. The LOFAR telescope implements a pipeline called \textsc{Cobalt} \citep{LOFAR:2018}, that uses similar design methodologies.

\subsection{Disk I/O}

Writing to disk drive (or more generally, a `data sink') is the final stage of most processing pipelines. The write speed of a drive is in general much slower than RAM, so this step can be a bottleneck. The two prevailing technologies are magnetic storage hard disk drives (HDDs) and solid state drives (SSDs) that use flash memory. HDDs have spinning ferromagnetic disk platters, with maximum write speeds in the range 80--160\,MB/s, for contiguous writes (for a current-generation 7200\,RPM HDD). As of writing, capacities of up to 16\,GB are commercially available.

SSDs have no moving components and offer higher write speeds (1--3\,GB/s), but are more expensive per TB of storage space, and have smaller overall storage capacity. The flash memory in SSDs can only be written a certain number of times before failure, meaning sustained writing to SSDs will result in a short (months) lifetime. When an SSD fails, large portions of data will be lost. In contrast, HDDs are more susceptible to damage from physical shock, but unlike SSDs early warning signs are often given before failure, and total data loss is less common.

Data loss due to drive failure can be mitigated by using RAID (Redundant Array of Inexpensive Disks) configurations to introduce data redundancy. In a RAID configuration, multiple physical drives are combined into one virtual drive, and data are distributed across disks. Data read and write speeds can also be increased significantly, depending on the RAID configuration and number of drives.

On multi-server installations, distributed filesystems like Lustre and Ceph\footnote{\url{https://ceph.io/}} are often used, so that all servers can access shared data storage via Ethernet. In this case, data may not be local to the compute resource (known as data locality); care must be taken that access patterns and read/write speeds can be sustained. 

\subsection{Performance tuning}

Getting optimal performance out of a compute server requires knowledge of the underlying hardware, and is a vast topic. Here, we briefly detail a few key points to consider when implementing a real-time stream processing pipeline, that may not be commonly encountered in general CPU/GPU code optimization.

\paragraph{NUMA awareness.} The components of a compute server are connected via buses, each with a finite bandwidth. Significant bottlenecks can arise on these buses if poor data transfer patterns are used. An important consideration for computers with multiple CPUs is that each CPU has its own system bus, and is the centre of what is known as a non-uniform memory access (NUMA) node. Things like NICs, GPUs, and RAM are associated with a particular NUMA node, depending on which system bus they are located on. It is possible, but undesirable, for a NIC to be located on a different NUMA node than a GPU card. This is undesirable as memory access across NUMA nodes id slower than access within the node.  

 The bandwidth between NUMA nodes is determined by the interconnect technology. Intel chipsets use UltraPath Interconnect (UPI, previously QuickPath), and AMD chipsets use Infinity Fabric (previously HyperTransport). 

\paragraph{PCIe.} Another bottleneck concern is the PCIe (Peripheral Component Interconnect Express) bus. All peripherals, such as NIC and GPU cards, are connected via the PCIE bus, which has a limited number of lanes. As of writing, a PCIe v3.0 slot for a graphics card has 16 lanes (x16), each with 1\,GB/s bandwidth. Most NICs are housed on x8 slots, which means a maximum of 8\,GB/s can be offloaded (64 Gb/s). The latest revision, PCIe 4.0, is not yet widespread but will double the bandwidth per lane.  

\paragraph{Memory bandwidth.} The memory bus can also be a bottleneck. For DDR4 memory (double data rate v4), the bandwidth is 256$\times$ the clock speed; for DDR4-2666 this is 68\,GB/s. 

\paragraph{IRQ core binding.} Interrupts (IRQs) are by default shared between CPU cores, but they are problematic for real-time stream processing. One can manually assign a CPU affinity to individual IRQs to stop interrupts being sent to active CPU cores. Several pipelines provide tools to aid in IRQ bindings. 

\paragraph{Kernel tuning.} The default parameters used in Linux kernel are not optimal for real-time processing and high-speed data capture \citep{LEDA:2015}. A specific suggestion is to increase the permitted buffer size for TCP in {\tt /etc/sysctl.conf} file ({\tt net.core.wmem.max} and {\tt net.core.rmem.max}). We recreate the suggestions from \citet{LEDA:2015} in Tab.\,\ref{tab:kernel-tuning}, but caution that these may not be optimal for a given system.  

\paragraph{Direct I/O.} In modern operating systems, the kernel may try to cache a write request in memory. Consecutive read requests can then be much faster, as the data are already loaded into memory. Direct I/O (using the {\tt O\_DIRECT} flag in C), forces the data to be written directly to the disk (in 512\,B chunks). For sustained high-bandwidth disk writes, direct I/O is preferable to avoid unnecessary caching. 

\begin{table}[]
\small
\begin{center}
\caption{Modifications to {\tt /etc/sysctl.conf}, used in \citet{LEDA:2015} to improve data transport performance. \label{tab:kernel-tuning}}
\begin{tabular}{rl}
Parameter & Value \\ \hline
kernel.shmmax & 68719476736 \\
kernel.shmall & 4294967296 \\
net.core.netdev max backlog &  250000 \\
net.core.wmem max & 536870912\\ 
net.core.rmem max & 536870912 \\
net.core.rmem default & 16777216 \\
net.core.wmem default & 16777216\\
net.core.optmem max & 16777216 \\
net.ipv4.tcp mem    & 16777216 16777216 16777216 \\
net.ipv4.tcp rmem   & 4096 87380 16777216 \\
net.ipv4.tcp wmem   & 4096 87380 16777216 \\
net.ipv4.tcp timestamps & 0 \\
net.ipv4.tcp sack   & 0 \\
net.ipv4.tcp low latency & 1 \\ \hline
\end{tabular}
\end{center}
\end{table}

\section{Discussion}\label{sec:discussion}

High-performance stream processing systems are fundamental to the operation of radio telescopes, and define the science data products that the telescope can provide. Building a high-performance system remains a challenging engineering task, but the shift toward industry-standard Ethernet is a welcome simplification. 

This chapter has covered a lot of ground, with the aim of giving a broad overview of the technologies and approaches used for real-time stream processing in radio astronomy. In Section\,\ref{sec:stream-processing} and Section\,\ref{sec:hetero-dsp} we introduced stream processing concepts and heterogeneous signal processing systems. Section\,\ref{sec:ethernet} gave an overview of Ethernet networking, which is used extensively in modern systems. We briefly introduced first-stage processing in Section\,\ref{sec:frontend}, and discussed data redistribution in Section\,\ref{sec:transport}. Second-stage processing, most commonly performed on server-class computers, was presented in Section\,\ref{sec:backend}. This chapter concludes with a future outlook and discussion of promising next-generation technologies.

\subsection{Future outlook}

As of writing, Intel has a $>$95\% share of the high-end server market, based on the June 2019 Top500 supercomputer list (\url{https://www.top500.org/}), and NVIDIA has a $>$90\% share of accelerator cards. However, AMD is expected to gain a significant market share, due to beating Intel to 7-nanometer process and other advances; this is well evidenced by the Frontier supercomputer, due for completion in 2021, which will be powered by AMD chips and is expected to be the World's fastest supercomputer\footnote{\url{https://www.amd.com/en/products/frontier}}.

In theory, OpenCL supports CPUs, GPUs and FPGAs, and FPGA accelerator PCIe cards are becoming more common. In practice, differences in architecture mean that code still needs to be written to target a given architecture for best results. Nevertheless, we may see OpenCL become more commonplace; the Heterogeneous Systems Architecture (HSA) foundation\footnote{\url{http://www.hsafoundation.com/}} provides a specification with similar goals. 

A new type of memory, marketed by Intel as Optane\textsuperscript{TM}, blurs the line between volatile RAM and flash storage, and products exist in both DDR and SSD formats. This, and similar technologies like Micron's 3D XPoint\textsuperscript{TM}, may prove useful for buffering large volumes of data, avoiding disk I/O bottlenecks. We may also see adoption of RDMA access to storage devices, by approaches such as the NVMe over fabric (NVMe-oF) specification \footnote{\url{https://www.nvmexpress.org/}}. 

We expect to see 25 GbE/100 GbE become commonplace in the coming years. Adoption of IP version 6 will also continue to increase, which brings the option of larger packets. IPv6 defines an optional jumbo payload, allowing single packets of several gigabytes, which would alleviate issues capturing high packet rates and allow larger data stream frames to be transmitted in a single packet. So-called `smart' NICs, which have an FPGA on the NIC itself, may also become more prevalent.

With the huge momentum of deep learning and AI, we also expect to see machine-learning methods move into first or second-stage processing, to help form novel science data products. Work to interface frameworks such as tensorflow\footnote{\url{http://www.tensorflow.org}} with pipelines is an area worthy of investigation.

\section{Acknowledgements}\label{sec:acknowledgements}

D. Price thanks A. Jameson, B. Barsdell and D. Macmahon for their valuable insights into CPU/GPU processing over the years.

\bibliographystyle{apalike}
\bibliography{references.bib}

\end{document}